\documentclass[a4paper, amsfonts, amssymb, amsmath]{article}
\usepackage[utf8]{inputenc}
\usepackage{color}
\usepackage[dvipsnames]{xcolor}
\usepackage{xfrac}

\usepackage{appendix}
\usepackage{soul}
\usepackage{ulem}
\usepackage{graphicx}
\usepackage{soul}
\usepackage{authblk}
\usepackage{amsmath,amssymb,amsfonts,graphicx}

\usepackage[a4paper,top=2cm,bottom=2cm,left=2cm,right=2cm]{geometry}

\usepackage{tikz}
\usepackage[pdftex, pdftitle={Article}, pdfauthor={Author}]{hyperref}

\title{Glueballs at high temperature within the Hard-Wall holographic model}

\author[1]{Matteo Rinaldi 
 \footnote{Corresponding author email: matteo.rinaldi@pg.infn.it} }
    \affil{Dipartimento di Fisica e Geologia. Università degli studi di
 Perugia. INFN section of Perugia. Via A. Pascoli, Perugia, 06123, Italy.}

\author{Vicente Vento}

\affil{Departamento de F\'{\i}sica Te\'orica-IFIC, Universidad de Valencia- CSIC,
46100 Burjassot (Valencia), Spain.}

\begin{document}
\maketitle

 % Leave empty to omit a date

\begin{abstract}
In this investigation an holographic description of the deconfined phase
transition of
scalar and tensor
glueballs  is
presented
 within the so called hard-wall model. The spectra of these bound states of
gluons have been calculated
from the linearized Einstein equations for a graviton propagating
from a thermal $AdS_5$ space to an AdS Black-Hole. In this framework,
the deconfined phase is reached via a two steps mechanism. 
We propose {that}
the transition between the AdS thermal sector to the BH is described via
a  first order phase transition, with discontinuous masses at the
 critical  temperature,
which has been determined by Herzog's method of regulating the free
energy densities.
 {Then,}
the glueball masses diverge with increasing  $T$ in the
BH phase and
thus lead to deconfined states \`a la Hagedorn.

\end{abstract}

%\keywords{meson, glueball, gravity}

\maketitle

\section{Introduction}

A successful strategy for applying the  AdS/CFT correspondence and holography
\cite{Maldacena:1997re,Witten:1998zw}
to hadron physics is the so-called
bottom-up  approach. In this framework, one  starts from some non perturbative
features of QCD and
attempts to construct its five-dimensional holographic dual. One implements
duality in nearly conformal conditions defining QCD on the four dimensional 
boundary and introducing a bulk space which is a slice of $AdS_5$ whose size is 
related to $z_0 \sim 1/\Lambda_{QCD}$
\cite{Polchinski:2000uf,Brodsky:2003px,Erlich:2005qh,DaRold:2005mxj,
BoschiFilho:2005yh}. This is
the so called hard-wall (HW) approximation. Later on, in order to reproduce the
Regge
trajectories of the hadronic spectrum, the  so called soft-wall model
was introduced
\cite{Karch:2006pv,FolcoCapossoli:2015jnm}.
Within the bottom-up strategy and in both, hard-wall
and the soft-wall approaches, glueballs 
arising from the correspondence 
of  fields in $AdS_5$ have  been studied 
\cite{BoschiFilho:2005xct,Colangelo:2007pt,Forkel:2007ru,
Li:2013pta,Li:2013oda}. 
However,
we have recently proposed the calculations of the spectrum of the scalar and
tensor glueballs under the
assumption that in this holographic approach,
the dual operator to the glueballs
could be  the
graviton, the latter thus plays a significant role to describe the lowest
lying
glueballs. We have studied the problem in hard and the graviton soft-wall
models and found an excellent description of the data with very few parameters
\cite{Rinaldi:2017wdn,Rinaldi:2018yhf,Rinaldi:2020ssz,Rinaldi:2021dxh}.
The main result of these investigations is that
 we do not need to introduce
additional fields into any $AdS_5$ to describe the glueballs, the gravitons indeed
satisfy the duality boundary conditions and are able to describe the elementary
scalar and tensor glueball spectra.
{Due to the exploratory nature of the}
the present investigation,
the HW  AdS/QCD model
 has been {used to} study the
 deconfinement phase transition.
Since we have proposed
that the scalar and tensor glueball spectrum is associated to the graviton of
the theory \cite{Rinaldi:2017wdn,Rinaldi:2021dxh}, it is therefore natural to
generalize this association to the graviton propagating in a black-hole (BH)
space.   Thus, in the following we have studied
the graviton spectrum when a BH background is considered in order
to describe the mass dependence on the temperature of the environment and
compare the new result with the previous calculations.
We recall that  much research has been carried out to determine
the deconfinement temperature and the behaviour of the glueball and meson
spectra after the phase transition
\cite{Kajantie:2006hv,Colangelo:2009ra,Braga:2017apr}.

In the present analysis,
we found out that the deconfinement phase is reached via a two steps mechanism.
We propose  {a strategy }
 to describe the transition from the
AdS thermal phase, i.e. the low temperature region, to the BH sector, i.e. the
high
temperature sector. 
{In particular}, the Hawking-Phase phase transition is a first order
phase transition
 at the temperature obtained by  Herzog \cite{Herzog:2006ra}.
These calculations have
{been}
 extended to the excited states.

\section{Scalar and tensor glueballs at zero temperature}
\label{section2}

In this investigation, we consider the holographic description of glueball
states via the hard-wall model.
Virtues and
inconveniences of this model have been thoroughly discussed, e.g., in Refs.
\cite{Polchinski:2000uf,Karch:2006pv,DaRold:2005mxj,Rinaldi:2017wdn}. Here we
start from
the gravity action:

\begin{align}
\label{action}
 \mathcal{I}  = \frac{1}{16 \pi G_5} \int d^5x~\sqrt{-\bar g} \;
[R + 2 \Lambda],
\end{align}
where $R$ is the Ricci curvature and $\Lambda$ the corresponding cosmological
constant. It can be easily shown that:
 
 \begin{align}
 ds^2= \frac{L^2}{z^2} (dt^2 +d \vec{x}^2 +dz^2)
 \label{AdSmetric}
 \end{align}
 is a solution of the Einstein-Hilbert equations if $\Lambda = -6/L^2$.
The equation of motion (EoM) for a graviton propagating in the thermal
$AdS_5$ space can be obtained from the Einstein equation for a perturbation
in this space.
By performing a linear expansion

 \begin{table} [htb]
\begin{center}
\begin{tabular} {|c c c c c c c|}
\hline k & 1 & 2 & 3 &4 &5& ...\\
\hline
D scalar $\,$ & 5.136 & 8.417 & 11.620 & 14.796 &17.960 &... \\
\hline
N scalar & 3.832 & 7.016 & 10.173 & 13.324& 16.471 &... \\
\hline
\end{tabular}
\caption{Energy modes for the scalar glueball with Dirichlet (D) and Neumann
(N) boundary conditions \cite{Rinaldi:2017wdn}.}
\label{modes}
\end{center}
\end{table}
 
 \begin{align}
 \bar{g}_{ab} = g_{ab} + h_{ab}
 \label{expansion}
 \end{align}
 one obtains the graviton equations of motion
 \cite{Constable:1999gb,Brower:2000rp}. We report here the result presented in
Ref. \cite{Rinaldi:2017wdn}
for
the scalar component of the graviton obtained from the standard $AdS_5$ metric,
i.e., at $T=0$:
 
 \begin{align}
 \frac{d ^2\phi(z)} {d z^2}  - \frac{3}{z} \frac{d \phi(z)}{d z} + M^2  \phi(z) = 0,
 \label{coldgraviton}
 \end{align}
 where $M$ is the mass of the scalar gravitons.
In the HW model, the confinement is realized by
restricting the maximum value
 of $z \leq z_0$ at which one imposes either Dirichlet or Neumann boundary
conditions.
The exact solution of the above equation has been shown in Ref.
\cite{Rinaldi:2017wdn} and the
 corresponding modes, in units of $1/L$, are shown in Table \ref{modes}.
As discussed in Ref. \cite{Rinaldi:2017wdn}, in the $AdS_5$ space,
 the scalar and tensor graviton equations are the same for the
HW model.

 \begin{table} [htb]
{\color{black}
\begin{center}
\begin{tabular} {|c |c| c| c |c |c |c|}
\hline
$J^{PC}$& $0^{++}$&$2^{++}$&$0^{++}$&$2^{++}$&$0^{++}$&$0^{++}$\\
\hline
MP & $1730 \pm 94$ & $2400 \pm122$ & $2670 \pm 222 $&  & &  \\
\hline
YC & $1719 \pm 94$ & $2390 \pm124$ &  &  &  &  \\
\hline
LTW & $1475 \pm 72$ & $2150 \pm 104$ & $2755 \pm 124$& $2880 \pm 
164 $& $3370
\pm 180$& $3990 \pm 277$  \\
\hline
SDTK  & $1865 \pm 25^{+10}_{-30} $  & & 
& & & \\
\hline
\end{tabular}  
\caption{Scalar glueball masses [MeV] from lattice calculations by MP
~\cite{Morningstar:1999rf}, YC~\cite{Chen:2005mg} and LTW
~\cite{Lucini:2004my} and the recent analysis SDTK~\cite{Sarantsev:2021ein} .}
\label{Gmasses}
\end{center}
}
\end{table}

\begin{figure}[htb]
\begin{center}
\includegraphics[scale= 0.8]{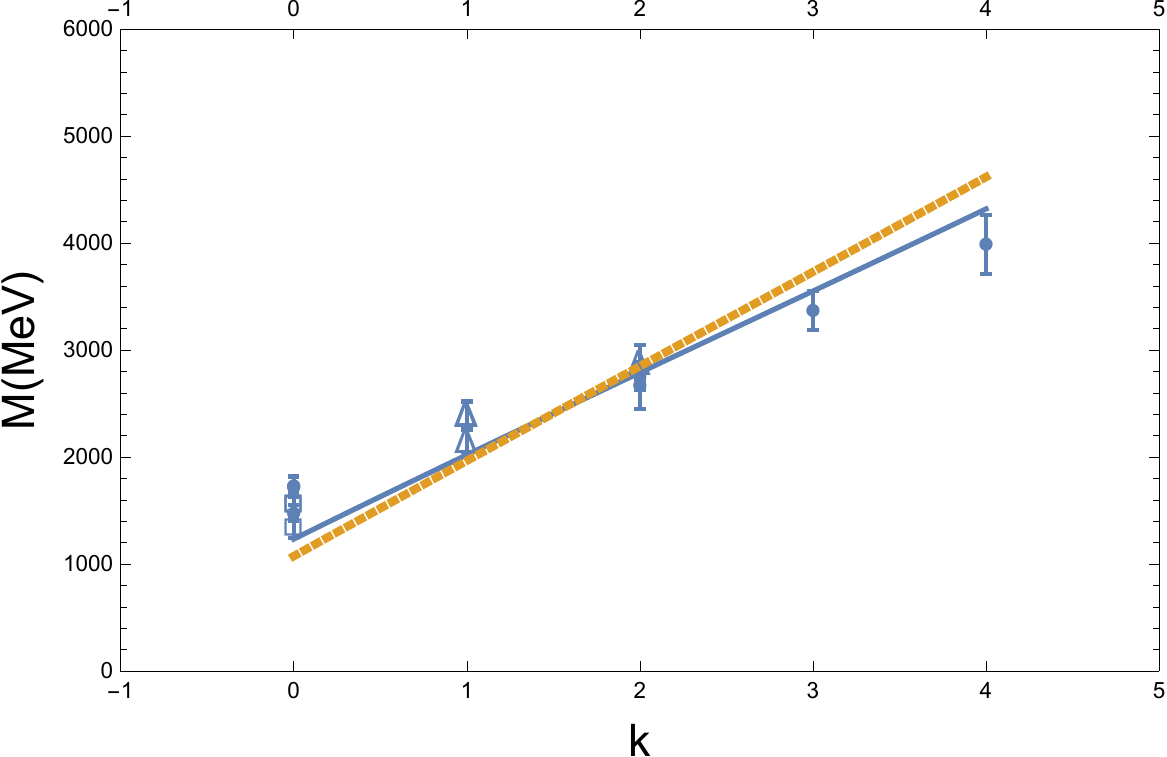}
\end{center}
\caption{Scalar and tensor glueball spectrum obtained within the hard-wall
model.
The solid lines correspond to
Dirichlet boundary conditions ($z_0=L_d^{-1} =250$ MeV) and the dashed lines
correspond to Neumann boundary
conditions ($z_0=L_n^{-1} =290$ MeV).
The full circles represent the scalar LQCD
masses, the squares the large N
limit scalar LQCD masses  and the triangles the tensor
 LQCD masses~\cite{Rinaldi:2017wdn}.}
\label{hardwall}
\end{figure}

Moreover, in Ref. \cite{Rinaldi:2017wdn},  the energy scale has been determined
by
fitting the lattice data
of scalar and tensor glueballs, shown in
Tab. \ref{Gmasses}. Results are also displayed  in Fig. \ref{hardwall}.
Let us remind that
in the hard-wall model the scale is given by $z_0=L$, i.e. the confinement
parameter.
The fit leads to $z_0 =L_d=1/250$ MeV$^{-1}$, for the
Dirichlet boundary conditions, while for the Neumann ones
$z_0 =L_N=1/290$ MeV$^{-1}$.

  { In closing this section, we compare the scalar and tensor
 glueball spectra obtained
from different holographic models. We report in Tab.} \ref{Gmmodels}
  { some numerical results. As one can see,
holographic approaches are powerful tools to explore the glueball spectra.
However, as discussed in, e.g. Refs.
\cite{FolcoCapossoli:2015jnm,Colangelo:2007pt,Capossoli:2015ywa},
 the standard Soft-Wall (SW) models,
where scalar fields dual to the glueball are considered,
cannot describe the overall spectrum, in particular they cannot reproduce
the ground and higher excited states at the same time. Therefore,
modifications, as those of Refs.
\cite{FolcoCapossoli:2015jnm,Colangelo:2007pt,Li:2013oda,Boschi-Filho:2012ijd,Capossoli:2015ywa},
 have been developed. However,
although the agreement with lattice data has been improved,
 with respect to the SW model, also in these cases,
a global overall good descriptions of scalar and tensor glueballs for low and
excited states has not been reached. We mention the model
of Ref. \cite{Brunner:2015oqa} which reproduces well the lowest modes
 if the first exotic state is skipped.
Among the modifications of the SW model we also recall the
recent GSW model
of Ref. \cite{Rinaldi:2017wdn} capable of describing the
overall scalar and tensor glueball
spectrum. Moreover, as already stated,
due to the exploratory nature of the present work, we focus on the
HW \cite{BoschiFilho:2005yh,Rodrigues:2016cdb}
model, which, despite its simplicity,  reproduces quite well the
 glueball spectra.}

\begin{table} [htb]
{\color{black}
\begin{center}
\begin{tabular} {|c |c| c| c |c |c |c|}
\hline
$J^{PC}$& $0^{++}$&$2^{++}$&$0^{++}$&$2^{++}$&$0^{++}$&$0^{++}$\\
\hline
GSW \cite{Rinaldi:2017wdn} & $1920$ & 2371 & $2830 $&2830  & 3289 & 3740 \\
\hline
SW1 \cite{Capossoli:2015ywa} & $2320 $ &3460  &2830  &  &3270  & 3660  \\
\hline
SW2 \cite{Capossoli:2015ywa} & $1840$ &4900  & $2610 $& &3190 &3690    \\
\hline
HW \cite{BoschiFilho:2005yh}  & $1630 $  & 2410 &2670
&3510 & 3690 & \\
\hline
Ref \cite{Brunner:2015oqa}  & $1487 $  & 2168 &2358
&3075 & 3202 & \\
\hline
\end{tabular}
\caption{ { Scalar and tensor glueball masses [MeV] from holographic models.
In the model related to the
 last line, corresponding to Ref. \cite{Brunner:2015oqa},
 the first exotic mode  has been skipped. } }
\label{Gmmodels}
\end{center}
}
\end{table}

\section{The glueball deconfinement phase transition}

In this section the glueball deconfinement phase transition mechanism, realised
within
the HW model, is described.
Within this approach, the first step, towards the deconfinement, corresponds to
a
Hawking-Page ~\cite{Hawking:1982dh} first order phase transition between the AdS
thermal space, see Eq. (\ref{AdSmetric}), at low temperature and an
asymptotically
AdS geometry containing a black hole

\begin{align}
ds^2 = \frac{L^2}{z^2} (f(z) dt^2 + d \vec{x}^2 + f^{-1} (z) dz^2),
\label{bhmetric}
\end{align}
at high temperature.  Here $f(z)= 1-{z^4}/{z_h^4}$, thus $z_h$ determines
the Hawking's temperature of the black hole solution $T_h = 1/(\pi z_h)$.
The comparison between the free energy densities of both phases leads to
a critical temperature~\cite{Herzog:2006ra}:

\begin{align}
\label{tem}
T_c = \frac{2^{\sfrac{1}{4}}}{\pi z_0}
\end{align}
where again $z_0$ is the infrared cut-off determining confinement
 and  the phase transition is characterized  by the relation $z_0^4 = 2z_h^4$
\cite{Herzog:2006ra}.
 Thus as the
temperature increases, the AdS thermal becomes unstable and the black-hole
becomes
stable. At $T_c$, the BH horizon forms inside the AdS cavity, between
the boundary and the infrared cut-off , at a radius $z_h< z_0$.
Here, from Eq. (\ref{tem}), the temperature can be determined from
$z_0$. In the present analysis we consider $z_0$ obtained from the previous
fit of the scalar and tensor glueball spectra. Numerically,
 we obtain
 $T_c= 95 $ MeV for Dirichlet boundary conditions, and $T_c= 110 $ MeV for
Neumann boundary conditions. In Tab. \ref{tcdata}, some lattice data and
previous
calculations of  $T_c$ are reported for comparisons.
As one can see, the above results are distant from those in Tab. \ref{tcdata}.
Nevertheless, if one  evaluates $T_c$, from Eq. (\ref{tem}), by fixing $z_0$
according
to the average of lowest glueball mass, see Table \ref{Gmasses}, then a more
realistic value could be found, i.e.:
 $T_c \sim 125 $ MeV for Dirichlet boundary
conditions and $T_c \sim 165 $ MeV for Neumann boundary conditions.
However, by comparing the above results with those in Tab. \ref{tcdata}, one
can realise that apparently the HW model needs further improvements, since data
related to gluodynamics lead to higher critical temperatures.
In the future we could consider other models, such those of Refs.
\cite{Colangelo:2007pt,Li:2013oda,Capossoli:2015ywa,Bernardini:2016qit,
Rinaldi:2017wdn,Rinaldi:2021dxh}, to calculate $T_c$.
 Other authors have used experimental values of
the meson spectroscopy
 to fit the deconfinement temperature. In the hard-wall model
the result is also too low $T_c \sim 125$ MeV, however they obtain higher values
for the soft wall model $T_c > 160 $ MeV
\cite{Herzog:2006ra,Kajantie:2006hv,Colangelo:2009ra,Afonin:2014jha}.

\begin{table} [htb]
\begin{center}
\begin{tabular} {|c | c |c|}
\hline
Reference & $T_c$ [MeV] & Features  \\
\hline
\cite{Herzog:2006ra} &122 & Hard-Wall with $z_0=1/323$ MeV
\\
\hline
 \cite{Herzog:2006ra} & 191 & Soft-Wall
\\
\hline
\cite{Cheng:2006qk} &$192^{+7}_{-4} $ &Lattice with quark dynamics
\\
\hline
\cite{Aoki:2006br}& $150 \pm 3$ &Lattice with quark dynamics
\\
\hline
\cite{Bazavov:2011nk}  &$154 \pm 9$ &Lattice with quark dynamics
\\
\hline
 \cite{Bernard:2004je}  & $169^{+12}_{-4}$&Lattice with quark dynamics
\\
\hline
\cite{Borsanyi:2010bp}  & 150-170 & Lattice data with quark dynamics
\\
\hline
\cite{Lucini:2012wq} & 250& Yang-Mills and large $N_C$ limit
\\
\hline
\cite{Boyd:1996bx} & $264 \pm 1$ & SU(3)
\\
\hline
\cite{Iwasaki:1996ca}& $276^{+3}_{-2}$ &  SU(3)
\\
\hline
\cite{Afonin:2014jha} & 250-270 & Soft-Wall and Improved Soft-Wall
\\
\hline
\end{tabular}
\caption{Some theoretical estimates of the critical temperature $T_c$.}
\label{tcdata}
\end{center}
\end{table}

\section{Scalar and tensor glueballs beyond the critical temperature}

In this section, we present the mode equations and solutions for scalar and
tensor glueball states dual to gravitons propagating in both the thermal
$AdS_5$ and black-hole spaces. We expect that
beyond the critical temperature the BH horizon forms inside the
AdS cavity between the boundary and $z_0$,  $z_0>z_h$. We now have to construct
the equations of motion
 for the gravitons with the black hole metric,
Eq.(\ref{bhmetric}). { To this aim,
use has been made of the procedure discussed above
for $QCD_3$. Then the EoM for the scalar graviton reads}
{
\begin{align}
\frac{d^2\phi(z)}{dz^2} + \left(\frac{2}{z} - \frac{5 z_h^4-z^4}{z (z_h^4-z^4)}\right) \frac{d \phi(z)}{dz} + \left(\frac{M^2  z_h^4}{z_h^4-z^4}+ \frac{ 256  z^6 z_h^4}{(z_h^4-z^4)(6 z_h^4-2z^4)^2}\right) \phi(z) = 0.
\label{scalar}
\end{align}}
{
and for the scalar graviton becomes}

{
\begin{align}
\frac{d^2\phi(z)}{dz^2} + \left(\frac{2}{z} - \frac{5 z_h^4-z^4}{z
 (z_h^4-z^4)}\right) \frac{d \phi(z)}{dz} + \frac{M^2  z_h^4}{z_h^4-z^4} \phi(z)
= 0,
\label{tensor}
\end{align}}
{
Furthermore, the tensor graviton EoM is the same as  that of an
external scalar field in the BH space~\cite{Constable:1999gb,Brower:2000rp}. 
One should notice that, at variance with the mode equations obtained in the
$AdS_5$ sector \cite{Rinaldi:2017wdn},
 in this case the scalar graviton has a different
mode equation with respect to that for the tensor graviton and external
 scalar field due to an additional
potential
term.} 

{
For the sake of simplicity,
 a constant $\lambda$ is introduced in front of the additional potential term
in Eq. (\ref{scalar})
 so that:
 $\lambda = 1$ corresponds to the scalar graviton EoM and $\lambda=0$ corresponds to
tensor graviton and external field EoM:}

\begin{align}
\frac{d^2\phi(z)}{dz^2} + \left(\frac{2}{z} - \frac{5 z_h^4-z^4}{z
(z_h^4-z^4)}\right) \frac{d \phi(z)}{dz} + \left(\frac{M^2  z_h^4}{z_h^4-z^4}+
\lambda\frac{ 256  z^6 z_h^4}{(z_h^4-z^4)(6 z_h^4-2z^4)^2}\right) \phi(z) = 0.
\label{scalar2}
\end{align}

\subsection{Solutions to the equation of motion in the BH background}

 As one might expect, Eq. (\ref{scalar2})  requires a delicate numerical
analysis. In order to  study the temperature dependence,  a useful change of
variable $w=z/z_h$  leads to

\begin{align}
\frac{d^2\phi(w)}{dw^2} + \left(\frac{2}{w} - \frac{5  -w^4}{w (1 - w^4)}\right) \frac{d \phi(w)}{dw} + \left(\frac{\tilde{\mu}^2  }{1 - w^4}- \lambda \frac{ 256 w^6}{(1- w^4)(6 -2 w^4)^2}\right) \phi(w) = 0,
\label{bhzhmodes}
\end{align}
where the quantity $\mu = M z_h$ is introduced. In order to find the modes of
this equation, one needs to integrate it  from
$w=0$ towards the horizon.
It is therefore useful to study the
 the behavior at $w=z=0$:

\begin{align}
\phi(w) \sim A w^4 + B
\end{align}
where $A,B$ are integration constants. For simplicity one can set, without
loosing
generality, $A=1$
and $B=0$ fixing thus the outgoing solution. The changing of the values of $A$
and $B$ leads
only to a modification of the shape of the mode function keeping
the energy modes fixed. This feature will be  explicitly shown later on after
the
resolution of the equation.

\begin{figure}[htb]
\begin{center}
\includegraphics[scale= 1.1]{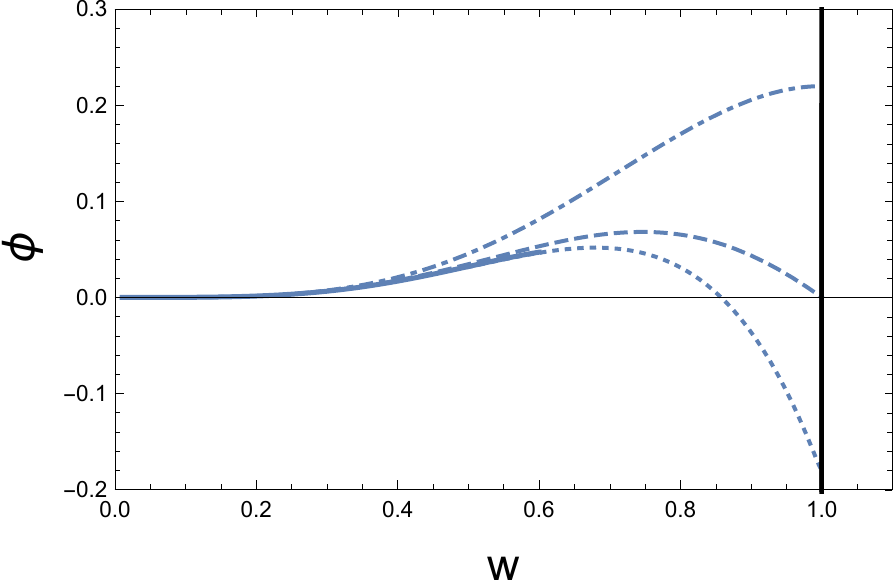} 
\end{center}
\caption{
The scalar glueball mode function
as a function of $w=z/z_h$.
The dashed curve corresponds to Dirichlet
 (d)
boundary conditions with mode value $\mu_d = 5.136$
 in the AdS thermal sector and for $z_0=L=z_h$. Dot-dashed line, same of the
dashed one, but
for the Neumann (n) boundary conditions and $\mu_n = 3.832$.
 The
solid-dotted curve corresponds to the black-hole AdS graviton solution. The
solid part of the latter curve is related  to the outward
solution and the dotted curve to the inward one. At matching  the mode
value is $\mu=5.487$ and the intercept with the black-hole radius occurs for
$a_0= -0.179$. }
\label{integration}
\end{figure}

Further, the equation must be the integrated from the horizon inward and
then one needs  to match the outward and inward solutions and to determine the
value of the energy mode. In order to study the behaviour of the solution close
to the horizon, needed for the numerical integration,
another change of variable is useful:
  $ v =1 - w^4$.   Equation (\ref{bhzhmodes}) now becomes:

\begin{align}
\frac{d^2\phi(v)}{dv^2} + \frac{1}{v} \frac{d \phi(v)}{dv} +
\frac{1}{16 v (1-v)^{\sfrac{3}{2}}}\left(\mu^2  + \lambda \frac{64 (1-v)}{(2 -
v)^2}\right) \phi(v) =0.
\label{bhhorizon}
\end{align}
where $v \rightarrow 0$ at the horizon, i.e. $ w \rightarrow 1$. The regular
solution at $v=0$  has the form of $\phi(v) = \sum_0^\infty a_n v^n$.
Substituting this ansatz into the equation and keeping only terms up to order 3,
one obtains  recurrence relations for $a_i$, with $i \geq 1$, the latter
functions of
the independent
$a_0$
coefficient. For the three first modes  one has:

\begin{align}
a_1 =&- \frac{(16 \lambda +\mu^2 )a_0 }{16}  \nonumber \\
a_2= &\frac{ (16 \lambda - 3 \mu^2 ) a_0 -(32 \lambda + 2 \mu^2 )
 a_1}{128 } \nonumber \\
a_3 =& -\frac{ (80 \lambda + 15 \mu^2 ) a_0 + (64 \lambda - 12 \mu^2 )  a_1
- (128  \lambda + 8 \mu^2 )  a_2}{1152 }.  \nonumber \\
\end{align}

\begin{figure}[htb]
\begin{center}
\includegraphics[scale= 1.1]{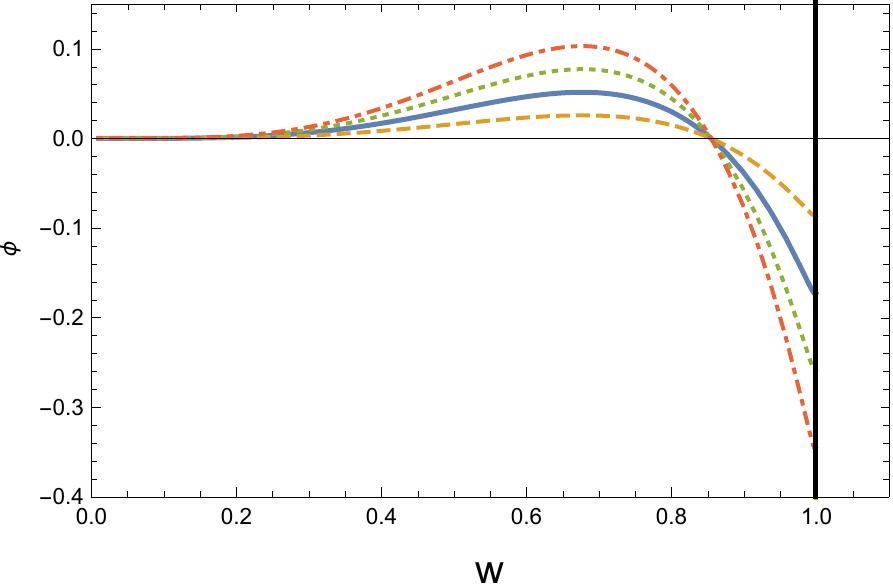} 
\end{center}
\caption{The solution of the differential equation (\ref{bhzhmodes}) for $
\mu= 5.487$ and
different values of $A = 0.5$ (dotted), $ 1$ (solid), $1.5$ (dashed), $2.0$
(dot-dashed) and $B=0$. The mode functions keep the same mass mode value but
with different
intercepts
$a_0= -0.091, -0.179, .0.276, -0.370$. }
\label{Aindependence}
\end{figure}

The approximate solution with the four first terms and its derivative
is used as initial condition for the numerical program at $v$ close to zero.
In Fig. \ref{integration}  the AdS thermal solutions for the Dirichlet
and Neumann boundary conditions, whose mode values are  $\mu_d =5.136 $ and
$\mu_n=3.832$ respectively, are shown for $z_0 = z_h$.
Let us remark that this choice is just an example. In addition,
also   the
AdS Black-Hole solution, obtained by matching  the outward and inward solutions
for  $\lambda = 1$ (corresponding to the scalar glueball), is displayed. The
matching numerically occurs for $\mu =5.487$ and $a_0 = -0.179$.
These two parameters, fixed the by matching, determine uniquely the
solution.
For the moment being we focus our attention on the scalar glueball. The
tensor case will be addressed later on. In order to proceed to the
 study of the behaviour of the solution, beyond the phase transition, it is
important to discuss
 the $A, B$ independence of the mass modes. The $B$ case is
straight forward since the change of $B$  simply implies  a displacement of
$\phi$. The numerical  $A$ dependence is shown in Fig. \ref{Aindependence}
where  the mode function is displayed for $A= 0.5, 1, 1.5, 2.0$ for the same
value of $\mu =5.487$. As one can see,  the matching is found for different
values of $a_0= -0.091, -0.179, -0.271, -0.370$. From the figure, one should
realise that the energy modes are determined  by the zero of the
mode functions which is reached at the same $w$ for all values of $A$.

The next step is to
 study the energy mode values as a function of the energy scale given by the HW
model, i.e., $z_0=L$. In this way we will obtain the temperature dependence of
the modes from the BH radius.
To this aim, we introduce a new variable
$u=z/z_0$. The differential equation becomes:

\begin{align}
\frac{d^2\phi(u)}{du^2} + \left(\frac{2}{u} - \frac{5 u_h^4 -u^4}{u
(u_h^4-u^4)}\right) \frac{d \phi(u)}{du} + \left(\frac{\tilde{\mu}^2
u_h^4}{u_h^4-u^4} +
 \lambda \frac{ 256 u^6 u_h^4}{(u_h^4-u^4)(6u_h^4 -2 u^4)^2}\right) \phi(u) = 0.
\label{bhmodes}
\end{align}
where now $\tilde{\mu} = M z_0$ and $u_h=z_h/z_0$.
One of the main advantages of moving to the $w$ and $u$ variables is that
Eqs.
(\ref{bhzhmodes},\ref{bhmodes}) have no direct dependence on any dimensional
external parameter thus the solution, for the lowest mode, is unique and
can be simply obtained for $z_0=z_h$, i.e. $u_h=1$.

On the other hand side, in order
 to obtain  the intercept, $a_0$, at the BH
radius, it is necessary, again, to change variable: $\omega = u_h^4- u^4$.
When $z \sim z_h$ then $v \sim \omega$ and the solution behaves like:
 $\phi(z) \sim \sum_0^n
\tilde{a}_n(z_0 ^4\omega)^n= \sum_0^n a_n (z_h^4 v)^n $, therefore $\tilde{a}_n
z_0^{4n} = a_n z_h^{4n}$, thus

\begin{align}
\tilde{\mu} = &M z_0 = M z_h \frac{z_0}{z_h} = \mu /u_h = 5.487/u_h, \label{uh}\\
\tilde{a}_0 = &a_0 \frac{z_0^4}{z_h^4} = a_0 /u_h^4 = -0.179/u_h^4.
\end{align}
Changing from the $w$ variable to the $u$ variable one can obtain the
temperature dependence of the mode functions  $\tilde{\mu}$ as given by Eq.
(\ref{uh}).

\subsection{Hard-Wall phase transition \`a la Herzog \cite{Herzog:2006ra} }

\begin{figure}[htb]
\begin{center}
\includegraphics[scale= 0.9]{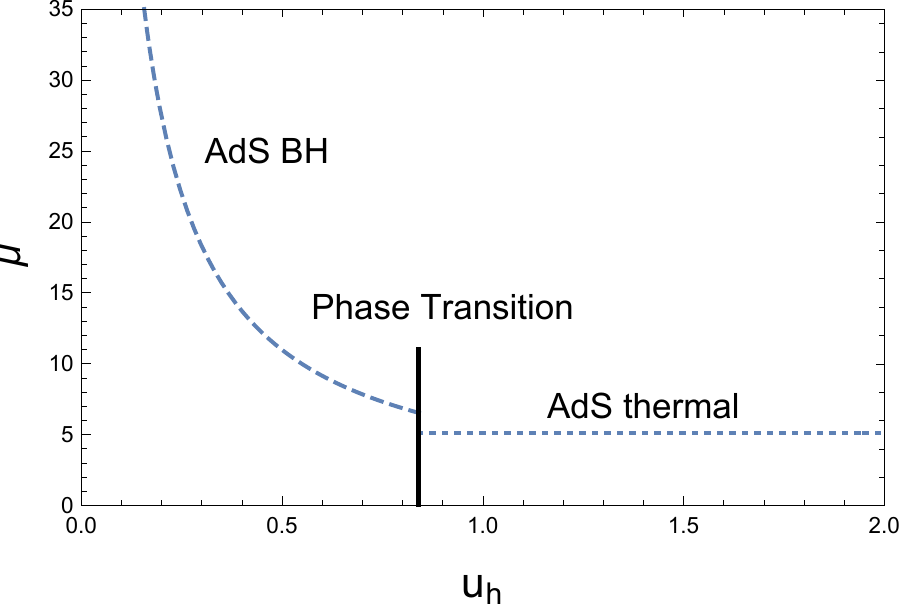}  \hspace{0.in}
 \includegraphics[scale= 0.9]{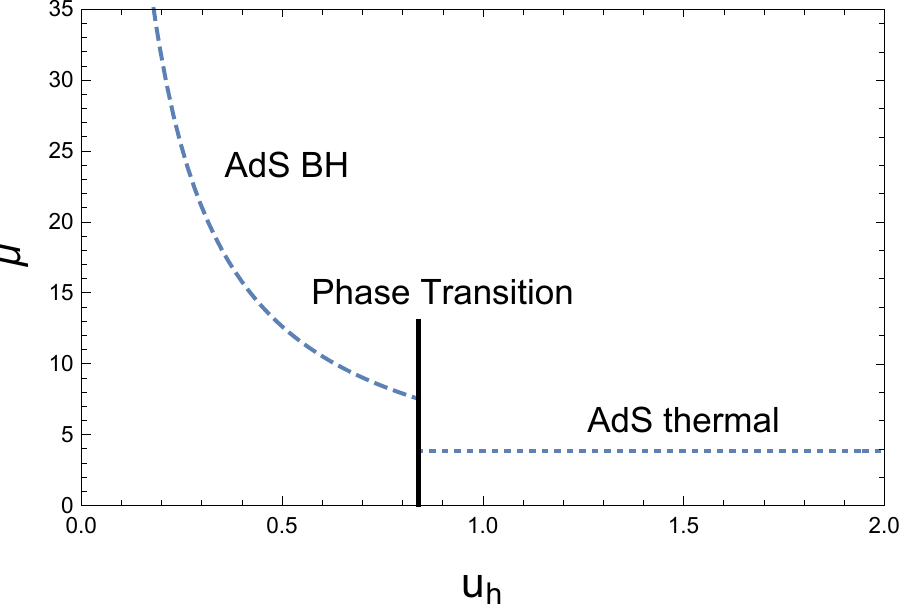}
\end{center}
\caption{The mass mode of the lowest scalar glueball state as a function of
$u_h$. The left
figure for the  Dirichlet boundary conditions, while the right one
for the Neumann boundary conditions. Here it is highlighted the behaviour
before and after
 the adimensional Herzog's critical temperature,
$u_h=2^{-\sfrac{1}{4}}$.  In
the AdS thermal phase we assume constant temperature dependence.
 }
\label{PhaseTransition1}
\end{figure}

\begin{figure}[htb]
\begin{center}
\includegraphics[scale= 0.9]{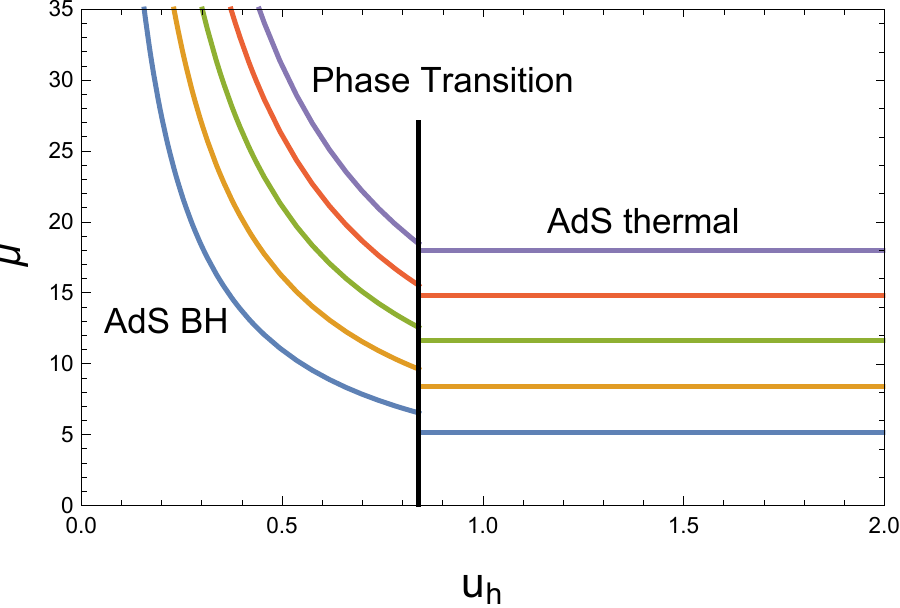} \hspace{0.in}
\includegraphics[scale= 0.9]{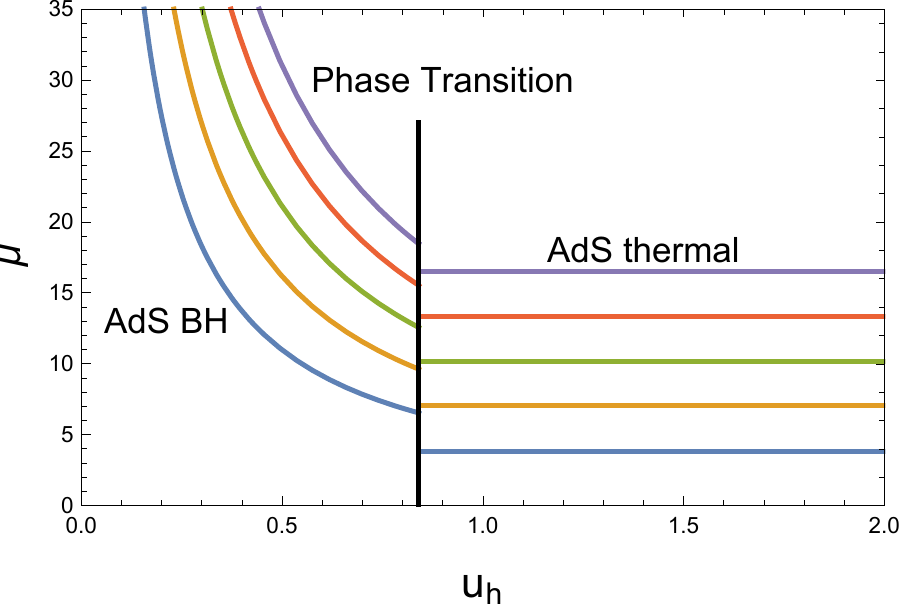}
\end{center}
\caption{The  spectrum of the scalar glueballs, in both sides of the
phase transition, for  the Dirichlet condition (left panel) and the Neumann
one (right panel). We are assuming here that in the AdS thermal phase the masses
of the
particles almost remain constant. }
\label{PhaseTransition2}
\end{figure}

Thanks to the above result,  it is now possible to describe the phase transition
from the AdS thermal to the AdS BH.
In Fig. \ref{PhaseTransition1}
we plot the AdS BH scalar glueball mass as a function of $u_h$, which is
basically the inverse of the temperature, and then we extrapolate the AdS
thermal mass at
$T=0$ ($u_h \rightarrow \infty$) toward $ u_h \rightarrow 0$, assuming a very
small temperature dependence in the
hard-wall model \cite{Colangelo:2009ra} stopping at Herzog's value
$u_h=2^{-\sfrac{1}{4}}$ \cite{Herzog:2006ra}. As one can see,
 there is a mass difference
at the boundary which is relatively large: $\Delta \mu_d =1.389 $,
corresponding to  347 MeV for the Dirichlet condition and $\Delta
\mu_n =2.693$, corresponding to 781 MeV, for the Neumann one.
Before closing this section,
let us now study the excited modes and how they behave at the phase transition.
To this aim, the higher modes have been calculated in the BH sector, which
determines
the high temperature dependence.
 The first five modes are shown in Table
\ref{BHmodes} for $u_h=1$. Moreover,
the BH modes are compared with those obtained in the AdS thermal sector,
at the Herzog's phase transition $u_h=2^{-\sfrac{1}{4} }$, in
Table \ref{scalarmodes}. The full dependence of the modes on $u_h$ is shown in
Fig. \ref{PhaseTransition2}.

\vskip 0.1 in

\begin{table} [htb]
\begin{center}
\begin{tabular} {|c c c c c c c|}
\hline k & 1 & 2 & 3 &4 &5& ...\\  
\hline
AdS BH scalar $\,$ & 5.487 & 8.081& 10.552 & 13.050 &15.511 &... \\ 
\hline
\end{tabular}
\caption{Energy modes for the scalar glueball  in the BH phase at $u_h=1$.}
\label{BHmodes}
\end{center}
\end{table}

\begin{table} [htb]
\begin{center}
\begin{tabular} {|c c c c c c c|}
\hline k & 1 & 2 & 3 &4 &5& ...\\  
\hline
AdS BH scalar $\,$ & 6.525 &9.610 & 12.548 & 15.519  &18.445   &... \\ 
\hline
AdS thermal  scalar  D& 5.136 & 8.417 & 11.620 & 14.796 &17.960 &... \\ 
\hline
AdS thermal  scalar N & 3.832 & 7.016 & 10.173 & 13.324& 16.471 &... \\ 
\hline
\end{tabular}
\caption{The energy modes for the scalar glueball AdS BH, at the phase
transition $u_h=2^{- \sfrac{1}{4} }$ and the AdS thermal modes at $T=0$.}
\label{scalarmodes}
\end{center}
\end{table}

\begin{figure}[htb]
\begin{center}
\includegraphics[scale= 1.1]{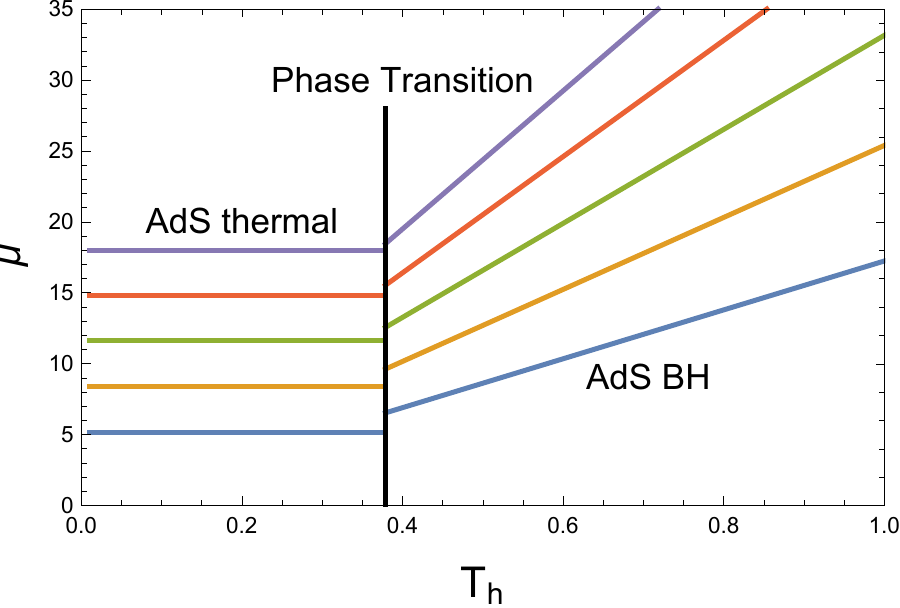}
\end{center}
\caption{Same of Fig. \ref{PhaseTransition2} but the modes are now functions of
the temperature in unit of the energy scale, i.e., $T={1}/(\pi u_h)$. }
\label{temperature}
\end{figure}

The results shown in Table \ref{scalarmodes} can be now converted into
physical glueball masses.
In terms of energy units, as already discussed, the fit of the glueball
spectrum leads to $z_0^{-1}=L_d^{-1} = 250$ MeV and $z_0^{-1}=L^{-1}_n = 290$
MeV, see Table
\ref{scalarmasses}.
One should notice that the mass differences at the phase transition is smaller
for the
Dirichlet solutions w.r.t.  the Neumann ones. Moreover this quantity
diminishes for the excitations. In closing,
the main approximation here assumed is the constancy of the dependence of
the modes with temperature in the AdS thermal phase. In this scenario, there
seems to be a first order phase transition, beyond the Herzog's temperature,
Eq.
(\ref{tem}), where
the deconfinement mechanism manifests itself in this  model by a high rise
in the masses of the states as the temperature increases, being  the masses
proportional to the latter, see for example Eq. (\ref{uh}). Furthermore,
following Hagedorn \cite{Hagedorn:1984hz}, at some
point the energy of the deconfined gluons will be smaller than that of the bound
states, and
 there the transition to the quark gluon plasma (QGP)
 will be reached.
Thus in this model, the transition from hadronic matter (HM)  to QGP  seems to
be a two step process, a  {first} order phase transition from hadronic matter to
highly massive glueballs (hadrons in general) and then a transition to QGP at
higher temperatures.
In some sense we are reminded of the scenario described by Shuryak
and Zahed \cite{Shuryak:2003ty} where one expects some glueball enhancement
mechanisms to appear \cite{Vento:2006wh}.
For the seek of clarity, the left panel of
Fig. \ref{PhaseTransition2} has been displayed again but showing directly the
energy modes as function of the temperature in units of ${1}/{z_0}$,
i.e. $T={1}/{(\pi u_h)}$,
see Fig. \ref{temperature}.
 The
importance of this figure lies in the AdS BH phase, which will repeat itself in
all other scenarios. The different excited states have divergent linear
trajectories, thus their masses separate more and more, leaving space for other
intermediate states, which could be associated to colour bound states, which our
colourless model does not contain, in line with the two step phase transition
already mentioned.

\begin{table} [htb]
\begin{center}
\begin{tabular} {|c c c c c c c|}
\hline k & 1 & 2 & 3 &4 &5& ...\\  
\hline
AdS BH scalar  D $\,$ & 1631 & 2402& 3137 & 3880  &  4611 &... \\ 
\hline
AdS thermal scalar D & 1284 & 2104& 2905 & 3699 &4490 &... \\ 
\hline
AdS BH scalar  N $\,$ &  1892& 2787 & 3639 & 4500  & 5349 & ... \\ 
\hline
AdS thermal scalar N & 1111 & 2034& 2950& 3863 & 4777 &... \\ 
\hline
\end{tabular}
\caption{The masses  in MeV units
 of the scalar AdS BH glueballs, at the phase transition and
the AdS thermal modes at $T=0$ with scales $L_d^{-1} =250$ MeV and
$L^{-1}_n = 290$ MeV.}
\label{scalarmasses}
\end{center}
\end{table}

 \vskip 0.1in

\subsection{Tensor glueballs  }
 
\begin{figure}[htb]
\begin{center}
\includegraphics[scale= 1.1]{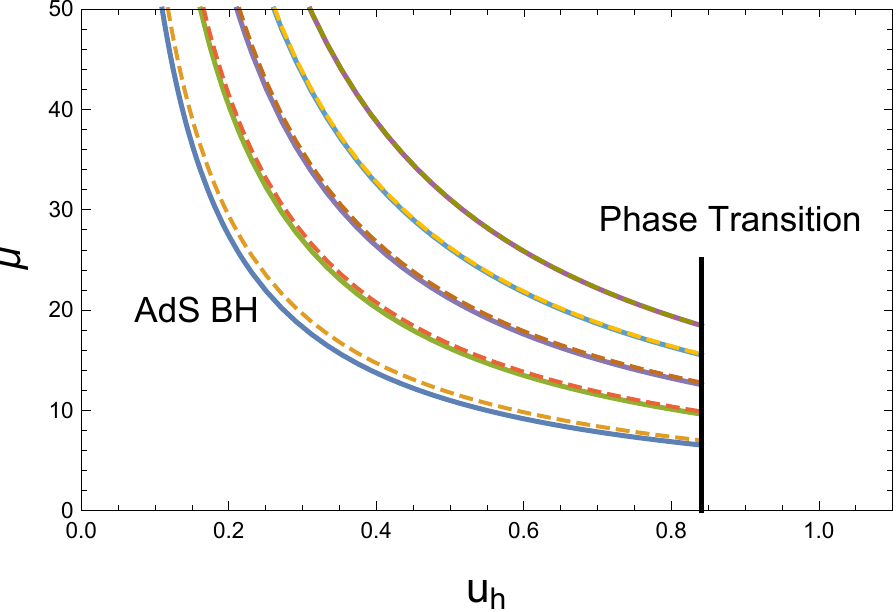} 
\end{center}
\caption{
The tensor glueball spectrum (dashed line) in the AdS BH sector, i.e., below
the horizon, as a function of $u_h$,
compared with
that of scalar glueballs (solid lines).
}
\label{excitationsT}
\end{figure}

Finally, let us now discuss the tensor components. We recall that the tensor
glueball states and the scalar glueballs are degenerate in this model in the
$AdS_5$. However, the mode equation in the BH AdS
background
 is different from that
of
the scalar graviton equation, see Eqs. (\ref{tensor},\ref{scalar}).
In Table
\ref{scalartensormasses} we
compare the masses of both the glueball components in the BH AdS sector
 at the phase transition and, in
Fig. \ref{excitationsT}, the temperature dependence of the the tensor
graviton spectrum (dashed)  and the scalar graviton spectrum (solid) are
respectively displayed.
As one can see,
the effect
of the additional term, proportional to $\lambda$ in Eq. (\ref{scalar2}), is
small. Moreover, the scalar modes
become lighter than the tensor ones as
expected. Therefore, we can conclude, that the deconfinement phase transition
mechanism for the
tensor glueballs follows the one already described for the
scalar case.
{Before closing this section, we can interpret the above
results in view of the HW model used. Indeed, as already stated
in sect. 4, the  external scalar field EoM in the BH background is the same
of that of the tensor component in the same space. Therefore,
 by comparing the spectra presented in this section
with those of the scalar graviton, e.g. see
Fig. \ref{excitationsT}, we prove that the modes of the scalar graviton
have lower masses than those of the scalar external field in the BH space. Such
a feature is consistent with the analysis of, e.g.
 Ref. \cite{Constable:1999gb}.}
 
 \begin{table} [htb]
\begin{center}
\begin{tabular} {|c c c c c c c|}
\hline k & 1 & 2 & 3 &4 &5& ...\\  
\hline
AdS BH scalar  D $\,$ & 1631 & 2402& 3137 & 3880  &  4611 &... \\ 
\hline
AdS BH tensor  D $\,$ &1747& 2467 &3150 &  3897& 4617 & ... \\ 
\hline
AdS BH scalar  N $\,$ &  1892& 2787 & 3639 & 4500  & 5349 & ... \\ 
\hline
AdS BH tensor  N$\,$ & 2027 & 2862 &3654 &4521  &  5356&... \\ 
\hline
\end{tabular}
\caption{The masses of the scalar and tensor  AdS BH glueballs, at the phase
transition  with scales $L_d^{-1} =250$ MeV and $L^{-1}_n = 290$ MeV.}
\label{scalartensormasses}
\end{center}
\end{table}

 \vskip 0.1in

\section{Conclusions}
\label{conclusions}
In this investigation the scalar and tensor glueball spectra have been
calculated within the holographic Hard-Wall model. In particular we have studied
the equation of motion for a graviton propagating first in a thermal AdS
space and then in a AdS Black-Hole in order to describe the glueball masses
as a function of the temperature. The scalar and tensor components are 
degenerate in AdS at $T=0$. However, when
the BH background is considered, such degeneracy is lost, and the tensor glueballs become heavier.
The energy scale of the HW model has been fixed by fitting lattice data of the
glueball spectrum.
Such a fit is quite good in particular when Dirichlet boundary conditions are
used.
Starting from these results, the mode energies of glueballs have been calculated
in both the AdS thermal and BH spaces. The outcomes of these evaluations have
been used to propose 
{a mechanism}
to describe the transition from the
AdS thermal sector to the BH one.
 {If} the masses do not depend strongly on the
temperature in the AdS thermal phase and at  Herzog's critical temperature
a first order Hawking-Page phase transition, between the low temperature AdS thermal phase and a high
temperature BH phase, takes place. 
 {Finally,}
the real transition to a purely deconfined state is described by a sharp rise
of the mass of the glueballs beyond
$T_c$.

The results of this investigation are quite dependent on the behaviour of the
modes in the AdS
thermal phase, but not so in the AdS BH phase, where the solutions of the
EoM are completely determined by the behaviour of the equations at the
horizon. We can conclude that
deconfinement
is realized following  Hagendorn's  mechanism  \cite{Hagedorn:1984hz},
consisting in a
rapid rise of the mass of the hadron states until they become heavier than a
system of unconfined gluons, forcing the glueballs to change from  bound states
to
unbound free particles. The resulting scenario  is  similar to that described
by Shuryak and Zahed \cite{Shuryak:2003ty} for a transition to an intermediate
phase of heavy colour bound states before the true deconfinement to QGP takes
over.

This investigation makes use of a specific model
and it should be {carried} out in more sophisticated models like the soft-wall
 or/and the
graviton soft-wall models. However, we  expect that a similar behaviour beyond
the horizon will take place there, with a rapid rising of the mode values
in the AdS BH phase.
In order to determine the type of phase transition, a temperature dependence
study of the modes  in the AdS thermal phase should be also  performed.

\section*{Acknowledgements}
The work
 was supported in part  by the MICINN and UE Feder
under contract FPA2016-77177-C2-1-P, by GVA PROMETEO/2021/083 and
by the European Union\'s Horizon 2020 research and innovation programme under
grant agreement STRONG - 2020 - No 824093.

\appendix
\section{Schr\"odinger solutions}

 It has been shown that Eq.(\ref{bhzhmodes}) can be transformed into a
Schr\"odinger type equations by  changing the function $\phi(z) = \beta(z)
\chi(z)$ where
$\beta{z}= z^2 z_h^2\sqrt{\frac{z_h-z}{z_h(z_h^4-z^4)}}$ followed by a change of
variable $z= z_h/(1+e^y)$  of the form

\begin{align}
- \chi''(y) = V(y) \,\chi(y),
\label{sch}
\end{align}
where

\begin{align}
V(y) = \frac{1}{4} +\frac{e^{2y} (15(1+ e^y)^8 -30(1+e^y )^4 -1}{4 (1+e^y)^2
((1+e^y)^4-1)^2} - \frac{\mu^2 e^{2y}}{(1-e^y)^4 -1} - \lambda \frac{64(1+e^y)^2
e^{2y}}{ (1+e^y)^4-1) (3(1+e^y)^4-1)^2},
\end{align}
where $y \in (-\infty, \infty)$ \cite{Constable:1999gb,Brower:2000rp}. We have
solved this differential equation numerically by starting from $-\infty$ towards
the right, and from $\infty$ towards the left and matching the solutions. In
order to do so we need the behavior of the potential which is

\begin{align}
\lim_{y \to -\infty} V(y)& \rightarrow \left(\frac{5}{4} -\frac{\mu^2}{4} -
4\lambda\right) e^y, \\
\lim_{y \to +\infty} V(y)& \rightarrow 4.
\end{align}
These limits tell us that the solution of the equation behaves at the limits as
\vskip 0.1in
\begin{figure}[htb]
\begin{center}
\includegraphics[scale= 1.1]{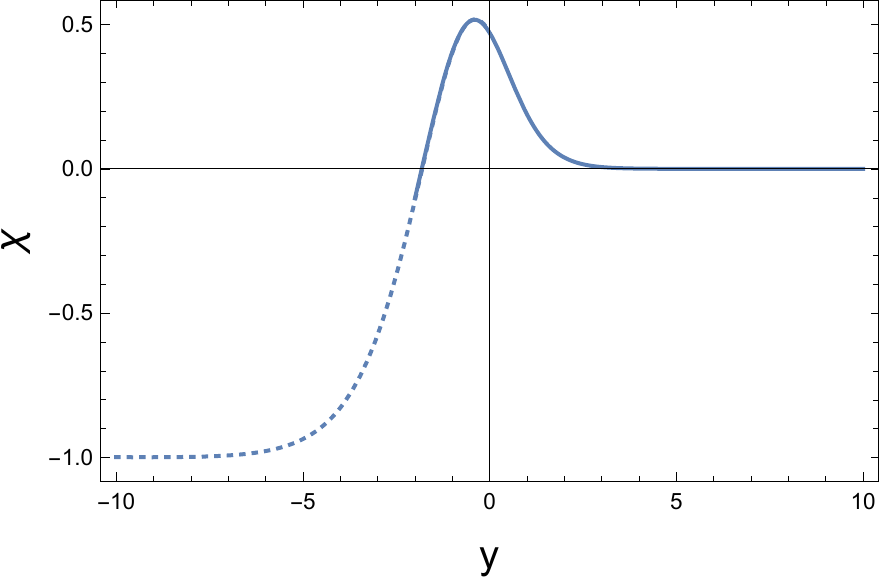} 
\end{center}
\caption{We show the  match between the left and right branches of the solution
to the Schr\"odinger equation Eq.(\ref{sch} for the lightest scalar glueball.}
\label{Schrodinger}
\end{figure}

\begin{align}
\lim_{y \to -\infty} \chi (y)& \rightarrow - e^{\delta e^y}, \\
\lim_{y \to +\infty} \chi(y) &\rightarrow \gamma e^{-\dfrac{2y}{\sqrt{\gamma}}}.
\end{align}
We fix $\delta = \frac{5}{4} -\frac{\mu^2}{4} - 4\lambda$ and we vary $\mu = M z_h$ and $\gamma$ to find the match. For $\gamma = 1.089$ we get the match shown in Fig. \ref{Schrodinger} exactly at $\mu= 5.51$ to be compared to $5.487$ for the method above. It is clear that this technique for solving the problem might be useful for the use of WKB methods but one looses the physical insight compared to our way of solving the problem, which establishes an direct comparison between the modes and mode functions in the AdS thermal phase and in the AdS BH phase.

\bibliographystyle{unsrt}

\bibliography{GravitonT.bib}

\end{document}